**The Evolution of Cryptography through Number Theory**


Fernando Peralta Castro

Moravian University[1]

May 4th, 2022



---

[1] This research was conducted while the author was affiliated with Northampton Community College. Current affiliation: Moravian University.




**Abstract**

Cryptography, derived from the Greek for "hidden writing," uses mathematical techniques to secure information by making it unreadable. Though the scientific study of cryptography is relatively recent, starting about 100 years ago, its development accelerated with the rise of the information age in the early 1900s. Cryptography has evolved from basic methods of protecting information to advanced systems using number theory to secure data for billions worldwide. The foundation of cryptography lies in mathematical concepts like modular arithmetic, the Euclidean algorithm, and Euler's totient function. This paper explores the progression of cryptography from early information-hiding methods, such as simple letter shifts, to modern algorithms grounded in advanced number theory. By analyzing historical and current practices, the paper illustrates how early cryptographic methods paved the way for contemporary encryption techniques. The evolution of cryptography demonstrates its critical role in safeguarding data in today's digital age.

## The Evolution of Cryptography through Number Theory

The term cryptography is derived from Greek and roughly translates to "hidden writing." Cryptography is the process of hiding information by utilizing various mathematical techniques to present the information in an unreadable format. Many historians point to the use of encrypted clay tablets and hieroglyphics as some of the earliest forms of this secret code writing known as cryptography in Mesopotamia circa 1500 B.C and in Egypt circa 1900 B.C.E (Kessler & Phillips, 2020). The modern study of cryptography has evolved as a division of computer science that utilizes mathematical functions, specifically number theory, to encrypt information so that it cannot be recognized by unauthorized parties. This process usually consists of two steps, encryption, and decryption. It takes readable text, known as plain text, and transcribes it into a series of characters that can't be understood. The process to create this series of unrecognizable characters is known as encryption and is known as the ciphertext. This serves as a way of transmitting information in a way that only the intended person can read it. Decryption is the process that transforms the encrypted information back into its original format. This process of encryption and decryption has been adopted as a way to protect all of our data; it helps encrypt online transactions, purchases made with credit card electronic chips and banking information. One of the main dangers with these online transactions and exchanges of data is that private data is being sent through a public network, known as the internet, in order to reach the intended receiver such as the bank. The need to keep this information safe led to the creation of secure ways to hide important information through number theory. As stated by Ghosal (2021), *"Number theory is probably one of the most important areas of Mathematics used in Computer Science and the basics behind all of modern Cryptography"* (p.35). Some of the earliest well-known forms of cryptography that utilized elementary number theory were known as the



Caesar cipher and the Vigenere cipher. These ciphers involved a shifting of the letters in the alphabet to encrypt a message. These methods began and evolved as a method of hiding information; through the evolution of number theory, these methods changed from simple shift ciphers to block ciphers where data is encrypted in blocks through the use of a key like the Rivest–Shamir–Adleman (RSA) cipher. This paper investigates the correlation between the earliest forms of information hiding, before cryptography was established as an official field of study, and correlates it to the RSA cryptographic algorithm that utilizes advanced number theory to secure information. The main differences between the early forms of information hiding and its modern counterparts is the amount of data that is being encrypted. The RSA algorithm was introduced during the late 1900s and has become a way to encrypt the data of millions of people while the Caesar and Vigenere ciphers had a usage rate that was much lower. Although they vary in the amount of data that is being secured, these 3 methods of information hiding have commonalities in the concepts of encryption and decryption and all encounter downsides to the encryption methods due to faults in the number theory that was used to create them.

Number theory is a branch of mathematics devoted to the study of positive integers; these numbers can be written as the set of natural numbers $\mathbb{N}$ where $\mathbb{N} = \{0, 1, 2, 3, \ldots\}$. Integers include a combination of two sets of numbers; these sets include the values of prime and composite numbers which are utilized to lay the foundation for cryptography in the RSA algorithm. Prime numbers can be defined as a number greater than 1 with two factors, 1 and the number itself. A composite number is defined as a number that has a multiple of at least two numbers other than 1 and itself. Number theory examines mathematics in discrete sets, such as integers $\mathbb{N}$ which include the sets of numbers that are discussed in this paper. Some of the basic applications of number theory are connected to basic mathematical topics such as divisibility and



modular arithmetic which are utilized to set the foundation for some of the earliest forms of

cryptography. An example of an early form of cryptography that utilized these basic

mathematical topics is the Caesar cipher. According to Samaila & Pur (2013), "An example of

such a cipher is the Caesar cipher; named after Julius Caesar, who reportedly used it to encrypt

information during the Gallic Wars by shifting each letter in the message three positions to the

right" (p.27). The Caesar cipher is an example of a shift or substitution cipher; this method

encrypts data by replacing the original letters in the message with a set number of characters

ahead in the alphabet. In the example of the Caesar cipher, the chosen shift was 3 characters

ahead in the alphabet. This cipher is a perfect example of utilizing mathematics in an attempt to

safeguard valuable information. Let's suppose that the phrase 'CIPHER' is the plaintext that

needs to be encrypted. Table 1 showcases how the cipher was intended to work. The

transformation or substitution can be represented by the two sets of alphabets in Table 1. One is

the English alphabet, labeled as 'Plaintext' in Table 1 and the encrypted alphabet is labeled as

'Encrypted' in Table 1 where the English alphabet is shifted 3 positions ahead. In this example,

the initial phrase 'CIPHER' would be encrypted to 'FLSKHU' as shown in Table 1.  These

procedures for encryption and decryption in the Caesar cipher follow a mathematical model that

was centered around modular arithmetic. The equation for encryption (Amanie, 2020, as cited in

Ryabko & Fionov, 2004 and Victor, 2014) describes the mathematical model for the Caesar

cipher as seen in (1).

$$c = (m + k) \bmod n \qquad\qquad (1)$$

In the equation shown in (1), c denotes the ciphertext, m denotes the position of a letter in the

alphabet by its numerical equivalent, k denotes the encryption key or the number of letters

shifted ahead and n denotes the size of the alphabet. With this knowledge, we can express the



modular equation for the Caesar cipher as seen in (2) and follow with an example of how the

cipher is set to work.

$$c = (m + k) \bmod 26 \quad \text{where m} \in Z_{26} \tag{2}$$

Here, mod 26 represents the 26 letters in the alphabet where m is equal to the numeric value of

any of these 26 letters. For example, the letter C in 'CIPHER' has a value of 3 because it is the

third letter in the alphabet. Table 2 shows the encryption process of the word 'CIPHER'. As

shown in Table 2, the equation $c = m + 3 \pmod{26}$ is the modular equation for the Caesar cipher

where the 3 is the number of letters shifted to the right. Thus, this shows another example of how

the cipher calculates the encrypted word to become 'FLSKHU'. By a similar definition, the

decryption equation (Amanie, 2020, as cited in Ryabko & Fionov, 2004 and Victor, 2014),

describes the mathematical model for the Caesar cipher as seen in (3) and shown in Table 3.

$$m = (c - k) \bmod 26 \quad \text{where k} \in Z_{26} \tag{3}$$

The Caesar cipher requires little computing resources and uses basic modular arithmetic to shift,

encrypt and decrypt a message. As a continuation, the alphabet was regarded as a cycle by

Caesar, so the letter that follows Z is A (Samaila & Pur, 2013). This makes it so that if the value

of $(c - k)$ is greater than 26, the letters would revert to the start of the alphabet by the use of

modular arithmetic as there would be a remainder that is equal to the letter. As time passed, the

Caesar cipher became obsolete, due in part because there are only 26 letters in the English

alphabet. Although the shift number could be incremented to make deciphering the message

harder, the decryption process would be able to be solved using modular arithmetic. The Caesar

cipher served as a method of hiding information, but after some time the method for cracking the

cipher was found. 'Cracking' the cipher typically refers to the method for decrypting any given

cipher by a person who does not have the information needed to decrypt it.  One of these



methods for cracking the cipher was known as frequency analysis. According to Kartha, R. S., & Paul, V. (2018), "The most sophisticated technique for the cryptanalysis of monoalphabetic cipher is called frequency analysis… It is based on the language we used for encryption where certain letters and combinations of letters occur with varying frequencies" (p.125). Frequency analysis is used to predict the frequency of a certain character in the alphabet in order to be able to predict the ciphertext. As an example, the letter E is the most common letter in the English language so a hacker can use frequency analysis to relate the letter E to the text that they are trying to decipher. The most frequent letters in the English language are shown in Table 4. By analyzing the letters that are shown most frequently in an encoded message, the shift cipher can be broken by using frequency analysis to see what those letters are. This was shown to be the biggest fault in the cipher and it rendered the cipher useless as a way to hide information. As time progressed, new advanced ways of hiding information would be invented. These new methods were influenced in part by the faults in previous methods of information hiding so that the new methods did not suffer from the same mistakes.

The Vigenere cipher is a method similar to the Caesar cipher that utilized elementary number theory to hide information. The Vigenere cipher is an example of a polyalphabetic cipher, which refers to any cipher based on substitution that uses multiple substitutions of the alphabet (Hamilton & Yankosky, 2004). This method utilizes a secret 'shift' word and utilizes multiple shifts across the message to encrypt the information within the message. The Vigenere cipher is another example of how number theory has come to aid in the evolution of cryptography. As explained previously, the Caesar cipher became an inadequate way of hiding information as time passed. In contrast, the Vigenere Cipher utilizes the same number theory principles as the Caesar cipher but helps prevent frequency analysis as a way of decrypting the



information. The Vigenere cipher utilizes a series of different Caesar ciphers based on a particular keyword. According to Hamilton & Yankosky (2004), "One of the most famous, private-key cryptosystems is the Vigenere cipher…Encryption and decryption using the Vigenere cipher were originally described in terms of a table known as the Vigenere Square and a secret keyword" (p.19). As the evolution of number theory progressed, the cryptosystem of the Vigenere cipher became a way of hiding information that corrected the faults that were found in the Caesar cipher. Introduced in the late 1500s and named after Frenchman Blaise de Vigenere, the Vigenere cipher can be explained through the properties of modular arithmetic (Hamilton, 2004). Let's suppose that the phrase 'MATHISREALLYCOOL' is the plaintext that needs to be encrypted for this cipher and 'DISCRETE' is the keyword. Table 5 showcases how the cipher was intended to work. This example shows the encrypted phrase for the phrase 'MATHISREALLYCOOL'. Typically these ciphers include very long keywords but this serves as a simplification to develop the understanding of the Vigenere cipher. By a similar definition of the Caesar cipher, (Hamilton & Yankosky, 2004) explains the encryption equation for the Vigenere cipher as seen in (4) and showcased in Table 5.

$$c = \ (p + k) \bmod 26 \quad \text{where } k \in Z_{26} \tag{4}$$

In this equation,  p is the plaintext letters, c is the ciphertext letters, k is the number of letters in the alphabet/the letters of the key and 26 are the letters of the alphabet. One of the main differences between the Vigenere and Caesar ciphers is that the letter 'A' is equal to the value 0. In the Caesar cipher, A=1, B=2, and Z=26. Now, in the Vigenere cipher, A=0, B=1, and Z=25. Here, the key is 'DISCRETE' but will be elongated to 'DISCRETEDISCRETE' so that the key is repeated to match the length of the plaintext. In this example the plaintext 'MATHISREALLYCOOL' will be encrypted to become 'PILJZWKIDTDATSHP'. This



algorithm uses the Vigenere Square, shown in Figure 1, to encrypt and decrypt the Vigenere

cipher. For every letter, there is an equivalent number based on its place in the alphabet up to 26

letters. As shown in Figure 1, the Vigenere cipher has the alphabet written out 26 times in a

combination of rows and columns known as the Vigenere square. As shown in the first column

of Figure 1, each row of the cipher is shifted one letter to the right when compared to the row

above it. This serves as essentially 26 different Caesar ciphers with increasing shifts. Manually,

the plaintext can be encrypted by hand for this example since it is fairly short. We can use the

letters in the key to tell us which rows to look for in the mapping. For example, for the first letter

of 'MATHISREALLYCOOL' we would start at row M on the Vigenere square. We would then

go down to the first letter of the word 'DISCRETE' which would be column D. This would

return the enciphered value of 'P' as shown in Table 5. Essentially, the Vigenere cipher is using

caesar cipher principles and modular arithmetic to increase the difficulty of the cipher. By a

similar definition of the Vigenere encryption cipher, (Hamilton & Yankosky, 2004) explains the

decryption equation for the Vigenere cipher as seen in (5) and showcased in Table 6.

$$p = (c + (26 - k)) \bmod 26 \quad \text{where } k \in Z_{26} \tag{5}$$

As Hamilton & Yankosky (2004) continues to explain, "The main entries of the table correspond

to ciphertext letters…To use the Vigenere Cipher two correspondents must first agree upon a

secret keyword or key phrase" (p.21) The encryption and decryption process of the Vigenere

cipher is shown by Table 6 and helps to show the evolution of number theory in the practice of

cryptography. The main flaw of the Caesar cipher was its small scope of only 26 characters

which allowed for frequency analysis of the English language to be used to decipher the

information. In the Vigenere cipher, modular arithmetic is used to divide the alphabet into a set

of 26 Caesar ciphers, increasing the difficulty and security of the cipher. As a form of encryption



and decryption, the Vigenere cipher served as a more sophisticated way of encrypting data and involved a secret key of random length. In the Vigenere Square, all 26 possible Caesar ciphers are represented (one per row). The strength of the cipher is that it is not subject to frequency analysis as a way to crack the cipher because the same plaintext letter will not always be encrypted to the same ciphertext letter. This cipher also uses probability to show that the cipher has $26^k$ possible values for any given letter. In this instance K is the length of the keyword. If a keyword is 7 letters long, there are 8 billion different ways that the phrase could be created in the Vigenere square. Now, the weakness of this cipher comes from its repeated key. If someone is able to guess the length of the key, modular arithmetic can be used to crack the cipher. This is because once the length of the key word is known the ciphertext can be treated as a Caesar cipher and broken. Since we used a 16 letter phrase in our example, we had to repeat our key of 'DISCRETE' twice in order to get the key to match the phrase. If the phrase is long enough and if it's accompanied by a short key, there will inevitably be repeated letters or patterns in the key and ciphertext since the key has to be repeated to match the length of the phrase. On the other hand, if a 150 word phrase is used along with a 150 word random non repeated key, this will make it very hard for this message to be deciphered since there are no inherent flaws or repeated keys in the message. The Vigenere cipher is an example of how the Caesar cipher was improved to make the process of information hiding more secure, even though it also had some downsides related to the number theory used to create the cipher.

Cryptography has evolved to become necessary in today's digital world where almost everything is accessible through the internet such as education, shopping, banking and social media (Tan, C. M. S., Arada, G. P, 2021). With all this information being hosted online, the problem becomes finding a way to secure all of this information in an efficient manner. This is a



problem that Julius Caesar and Blaise de Vigenere did not encounter or have to workaround when they introduced their respective ciphers. In order to provide data security and privacy for all of this data, advanced mathematical theorems needed to be incorporated which vary vastly for the elementary number theory used in the Caesar and Vigenere Cipher. To introduce the next algorithm, the RSA algorithm began as a way to encrypt and decrypt information during the information age. One thing that was common in ciphers like the Vigenere cipher was that a public key was discussed and written down ahead of time. In the Caesar cipher, the public key served as the number of letter shifts whereas in the Vigenere Cipher it was the keyword. Now, the RSA algorithm includes 2 sets of keys, one for encryption and one for decryption. Kulkarni (2017) explains that, "RSA is designed by Ron Rivest, Adi Shamir, and Leonard Adleman in 1978…It is an asymmetric (public key) cryptosystem based on number theory, which is a block cipher system" (p.99). The RSA algorithm works by using two prime numbers to generate a public key, which is used for key encryption, and the private key, which is used for authentication (Kulkarni, 2017). The security of the RSA scheme is based on the level of difficulty related to the factoring of very large integers (Lagarias, 2017). The RSA algorithm works by using many different sub-fields of number theory including modular exponentiation and congruences, general divisors and greatest common divisor(GCD), the Euclidean algorithm, and Euler's Totient theorem to encrypt and decrypt data (Karunankitha et al, 2021). At first glance, the advancement of cryptography through number theory is evident as the mathematical theorems used to encrypt and safeguard data have come a long way from the simple shifting of the letters of the alphabet that was used in the Caesar cipher.  To start the talk on RSA algorithms we need to introduce some of the concepts of number theory that are necessary for understanding how the algorithm works. First, I'll introduce the process of the RSA algorithm involved in the



key generation process, explain the math behind the process and then give an example. As

explained by Kulkrani (2017) the key generation process of RSA consists of five steps:

1. Choose two large prime numbers (p and q)

2. Compute n = p * q

3. Calculate φ(n) = (p-1)*(q-1)

4. Choose an integer e such that 1 < e < φ(n)

    a. Ensure that gcd(e, φ(n)) = 1

    b. Ensure that e and φ(n) are coprime

5. Compute an integer d, such that $d = e^{-1} \bmod \varphi(n)$

As explained previously, prime numbers along with many other solid number theory theorems

play a huge role in how the RSA algorithm encrypts data and maintains the data integrity. To

start, primes are described as numbers that can only have factors of 1 and itself. Primes have

become essential in most of the modern cryptography systems and are specifically used for key

generation in the RSA algorithm. Another important math definition in the RSA algorithm is the

idea of coprime or relatively prime numbers. Coprime refers to numbers that only have 1 as their

shared common factor. For example, 18 has factors 1, 2, 3, 6, 9, 18, and 35 has factors 1, 5, 7,

and 35, These are considered co-prime numbers because their only shared factor is 1. The idea of

coprime can be further explained by introducing the idea of the greatest common divisor or

GCD. The greatest common divisor of two or more integers is the greatest integer that divides

each of the integers. Two numbers can also be shown to be coprime if their greatest common

divisor is 1 (Karunankitha et al, 2021). To continue with the previous example, we can show that

the GCD of 18 and 35 is 1. If gcd(18,35) = 1, we can say that 18 is coprime to 35 and 35 is

coprime to 18. Typically, algorithms like the RSA algorithm use prime numbers that are



hundreds of digits long to encrypt their information. For really large prime numbers, the greatest

common divisor can be calculated through the Euclidean algorithm. The Euclidean Algorithm

was developed by the Greek mathematician Euclid as a technique for quickly finding the GCD of

two integers (Karunankitha et al, 2021). For example, let's use the numbers 270 and 192 as an

example to show how the Euclidean algorithm can help determine the greatest common divisor

of two integers. The equation for the euclidean algorithm (Karunankitha et al, 2021) can be seen

in (6).

$$a = bq + r \qquad 0 <= \text{r} <= \text{b}$$

$$b = rq_1 + r_1 \qquad 0 <= \text{r}_1 <= \text{r} \qquad\qquad (6)$$

In this example,

$$270 = 1 * 192 + 78$$

$$192 = 2 * 78 + 36$$

$$78 = 2 * 36 + 6$$

$$36 = 6 * 6 + 0$$

Here, we have shown that the greatest common divisor of 270 and 192 is 6 as 6 is the greatest

divisor before the equation has a remainder of 0. This can also be shown as gcd(270,192) = 6

because the last positive remainder is 6. The following theorem that is a basic part of the RSA

algorithm is Euler's generalization of  Pierre de Fermat's theorem which came to be known as

the Euler totient function. This equation is used in the encryption process of the RSA algorithm.

According to Karunankitha et al (2021), "Euler's totient function was first proposed by Leonhard

Euler; as a function that counts the numbers that are lesser than n and had no other frequent

divisor with n other than 1 (i.e., they are co-prime with n)" (p.3215). Euler's totient theorem is

used to count positive integers up to a given integer n that are relatively prime to n. If n is the



product of two primes p and q (i.e., n = pq), then the equation for Euler's Totient Function

(Karunankitha et al, 2021) is written as shown in (7) where Φ is the symbol phi..

$$\Phi\ (n) = \Phi\ (p)\ \Phi\ (q) = (p-1)\ (q-1)$$

$$(7)$$

Euler's totient function is also known as Euler's phi function where the equation is $1 < e < \varphi(n)$

or for which the greatest common divisor is gcd(en, e) = 1 (Karunankitha et al, 2021). As we

utilized and introduced before, the greatest common divisor is utilized again to meet the range of

the Euler function. The final step of the key creation process in the RSA algorithm is finding the

value of d, a member of the private key, by using the extended euclidean algorithm and the

equation (Karunankitha et al, 2021) shown in (8).

$$d \equiv e^{-1} \bmod (p\text{-}1)(q\text{-}1) \tag{8}$$

Kulkarni's (2017) findings show the following:

> The basic principle behind RSA is the observation that it is practical to find three very
>
> large positive integers e, d and n such that with modular exponentiation for all m, and that
>
> even knowing e and n or even m it can be extremely difficult to find d (p.100).

 Now that the steps for creating a key have been discussed, the encryption equation can be

introduced (Karunankitha et al, 2021) as seen in (9).

$$C = M^e \bmod n \tag{9}$$

 In the equation given, c is the ciphertext, and m is the message text. As Kulkarni continues to

explain, the RSA operations can be decomposed into three steps: Key Generation, Encryption,

and Decryption (Kulkarni, 2017). The following serves as an example of the RSA encryption and

decryption algorithm. The RSA algorithm is based on the difficulty to factorize a large integer.

For the purpose of understanding, this example will include very small prime numbers but this



algorithm typically utilizes prime numbers that are hundreds of digits long. Let's say that the

sender wants to encrypt and send a message with the letter "B" as a test of the validity of the

algorithm to the recipient. Let p = 2 and q = 7 since 2 and 7 are two small prime numbers. To

obtain the first part of the public key, we can multiply the two primes to compute n = pq. N = pq

would be equal to n = 14. We then introduce Euler's totient function, (Karunankitha et al, 2021)

to calculate the value of Φ(n) = (p-1)(q-1) as seen in (10).

$$\Phi(n) = (p\text{-}1)(q\text{-}1) = (2\text{-}1)(7\text{-}1) = (1)(6) = 6 \tag{10}$$

Now, we need to find the values of e, which stands for encryption value and d which stands for

decryption value. We then need to choose an integer (e) where e is an integer that is not a factor

of n and satisfies the equation 1 < e < phi(n). For this problem, the e value can be calculated by

using gcd(e, phi(n)) = 1 which means that e has to be greater than 1 but less than phi(n) which is

6. When plugged into the function the remainder of gcd(5,6) = 1 which is true. Through

calculation, the possible values of e are 5 and the possible values of d are 5, 11, and 17. Now the

public key consists of the pair (5,14). The public key encrypts information and the private key is

only known by the originator of the data for decrypting information which includes the value of e

and the value of n. Now, the phrase 'B' is equivalent to 2 due to the numerical value of the letter

B in the alphabet. This begins to relate to the Caesar and Vigenere ciphers as the place of the

letter in the alphabet determines its value in the algorithm. Now the encryption equation is

determined by the equation C = $M^e$ mod n as described by (9) where C is the modular

congruence of $M^e$ mod n. Thus the encryption value becomes $2^5$ mod 14 where 2 is the value of

the letter 'B'. It continues that 5 is the value of the encryption value and 14 is the value of n.

Now the problem is worked through using the equation that was introduced in (9).

$$C = M^e \bmod n$$



$$C = 2^5 \bmod 14$$

$$C = 32 \bmod 14 = 4$$

Now the encrypted message is the number 4, which translates to the value of the letter D in the alphabet. The "shifting" of all of the letters in a cipher would conclude with an encrypted message. Now to decrypt the message, we use the private key (11,14) and the decryption equation (Karunankitha et al, 2021) which is seen in (11).

$$M = C^d \; (\bmod \; n) \qquad\qquad (11)$$

$$M = 4^{11} \; (\bmod \; 14)$$

$$M = 4{,}194{,}304 \bmod 14 = 2$$

Now we see that 4,194,304 mod 14 is equal to a number with the remainder of 2. This would return that 2 is the encrypted message and the value of 2 translates back to the letter B in the alphabet. In this example, the public key is (5,14) and the private key is (11,14). The extensive number theory behind the RSA algorithms helps to showcase the evolution of cryptography through number theory. When compared to shift ciphers found in the Caesar and Vigenere cipher, the RSA cipher utilizes modular arithmetic in addition to many new theorems centered around the creation and factorization of extremely large prime numbers. Eventhough, the RSA algorithm utilizes a lot more number theory components than the previous ciphers discussed, there are still some downsides to the RSA algorithm. Some of the downsides of the RSA cipher are its slow data transfer rate and long calculation time due to the large prime numbers involved With this said, there is still no proven way to crack the cipher if a large enough key is used. RSA has been adapted to create SSL, Secure Sockets Layer, which has been one of the ways to secure an internet connection. As of now, RSA provides a very secure way to transfer data but this means that the RSA algorithm can also be used for illicit purposes to encrypt data. In addition, as the



prime numbers get longer, there is no apparent increase in security compared to hte computational power that is used to generate the key. This makes the RSA algorithm very resource heavy as it takes a really long time to do operations on these really large prime numbers. New methods of encrypting data have been adopted to secure data due to the very slow processing time caused by using prime factorization to create the keys in RSA.

       As was shown in this paper, the vast extension of the realm of number theory has been used to help evolve the field of cryptography and help keep information secure. Some of the first signs of cryptography being used came as early as 1900 BCE. More theorems like the Euclidean algorithm and Euler's Totient Function were introduced circa 300 B.C and in the 1700s respectively. These theorems, among others, were utilized for the RSA algorithm during the late 1900s. The range of theorems used to evolve cryptography spans from the earliest forms of elementary number theory to the more modern number theory additions. As explained by the paper, the main purpose for the evolution of cryptography has been to keep our information safe and the progression of number theory has allowed for the evolution of cryptography. As we have progressed technologically as a society, the need for cryptography has grown immensely. As explained by Kessler & Phillips (2020), "With the commercialization of the Internet and the dawning of the World Wide Web in the early 1990s, the government realized that there were legitimate needs for public use of strong cryptography" (p.2-3). As stated previously, cryptography plays a role in our everyday lives and does so through the advancements in number theory that have allowed for secure cryptosystems to be created. Cryptography is used when we open our phones, make phone calls, open our email accounts, order food online, use our credit cards, watch television, and even when driving our cars. All of this information is protected by the number theory behind the systems. If it wasn't for the advancements in number theory, the



field of cryptography would not have been able to advance in the way that it has. There is important information that is created and used every day and that needs to be encrypted for our privacy. We are able to use our phones to play games, browse the internet, and video chat through the power of number theory and cryptography. Through the processes of cryptography advancing from simple shift ciphers to the RSA cipher, the process of encryption and decryption has led the way for modern-day cryptography. These processes do have perceivable and exploitable weaknesses when compared to the modern cryptography systems. As discussed earlier, the Caesar and Vigenere ciphers had frequency analysis rendered the cipher worthless after the system was found out. When compared to the RSA cipher, which has no perceivable weakness when a big enough prime number is used, the evolution of cryptography is evident. The many mathematicians that helped introduce these theorems, like Euler and Euclid, paved the way for the evolution of cryptography trough their contributions to number theory. Although the time periods covered in this paper ranged from the B.C with the Caesar cipher to the late 1900s and the RSA algorithm, the need and power of cryptography have been evident even in our earliest forms of civilizations. Methods like the Data Encryption Standard (DES) and the Advanced Encryption Standard (AES) are systems that are considered more secure than the RSA algorithm and were introduced in the 1970s and 2001 respectively. Further research is still required to continue to make these cryptography systems more secure through number theory.

**Table 1**

*Caesar Cipher Alphabet Letter Shift*

| Plaintext | a | b | c | d | e | f | g | h | i | j | k | l | m | n | o | p | q | r | s | t | u | v | w | x | y | z |
|---|---|---|---|---|---|---|---|---|---|---|---|---|---|---|---|---|---|---|---|---|---|---|---|---|---|---|
| Encrypted | d | e | f | g | h | i | j | k | l | m | n | o | p | q | r | s | t | u | v | w | x | y | z | a | b | c |

*Note.* From *The Evolution of Cryptography through Number Theory (p.3), by Dawson Shores,*

*2021. Copyright 2021 by Dawson Shores*



**Table 2**

*Caesar Cipher and the Encryption Equation*

| Plaintext | m | c = m + 3 (mod 26) | Value of C | Ciphertext |
|-----------|-----|----------------------|------------|------------|
| C | 3 | 3 + 3 mod 26 | 6 | F |
| I | 9 | 9 + 3 mod 26 | 12 | L |
| P | 16 | 16 + 3 mod 26 | 19 | S |
| H | 8 | 8 + 3 mod 26 | 11 | K |
| E | 5 | 5 + 3 mod 26 | 8 | H |
| R | 18 | 18 + 3 mod 26 | 21 | U |

*Note.* Adapted from *The Evolution of Cryptography through Number Theory (p.3), by Dawson Shores, 2021. Copyright 2021 by Dawson Shores.*



**Table 3**

*Caesar Cipher and the Decryption Equation*

| Ciphertext | m | m = c - 3 (mod 26) | Value of C | Plaintext |
| --- | --- | --- | --- | --- |
| F | 6 | 6 - 3 mod 26 | 3 | C |
| L | 12 | 12 - 3 mod 26 | 9 | I |
| S | 19 | 19 - 3 mod 26 | 16 | P |
| K | 11 | 11 - 3 mod 26 | 8 | H |
| H | 8 | 8 - 3 mod 26 | 5 | E |
| U | 21 | 21 - 3 mod 26 | 18 | R |

*Note.* Adapted from *The Evolution of Cryptography through Number Theory (p.4), by Dawson*

*Shores, 2021. Copyright 2021 by Dawson Shores.*



**Table 4**

*The Average Letter Distribution of English Text*

| Letter | Frequency | Proportion | Letter | Frequency | Proportion |
|--------|-----------|------------|--------|-----------|------------|
| E | 11.1607% | 56.88 | M | 3.0129% | 15.36 |
| A | 8.4966% | 43.31 | H | 3.0034% | 15.31 |
| R | 7.5809% | 38.64 | G | 2.4705% | 12.59 |
| I | 7.5448% | 38.45 | B | 2.0720% | 10.56 |
| O | 7.1635% | 36.51 | F | 1.8121% | 9.24 |
| T | 6.9509% | 35.43 | Y | 1.7779% | 9.06 |
| N | 6.6544% | 33.92 | W | 1.2899% | 6.57 |
| S | 5.7351% | 29.23 | K | 1.1016% | 5.61 |
| L | 5.4893% | 27.98 | V | 1.0074% | 5.13 |
| C | 4.5388% | 23.13 | X | 0.2902% | 1.49 |
| U | 3.6308% | 18.51 | Z | 0.2722% | 1.39 |
| D | 3.3844% | 17.25 | J | 0.1965% | 1 |
| P | 3.1671% | 16.14 | Q | 0.1962% | 1 |

*Note.* From *A Hybrid Encryption and Decryption Algorithm using Caesar and Vigenere Cipher (p.4), by Tan, C. M. S., Arada, G. P., Abad, A. C., & Magsino, E. R., 2021,* as cited in *Gladwin S J and P Lakshmi Gowthami* 2020. *Copyright 2021 by Tan, C. M. S., Arada, G. P., Abad, A. C., & Magsino, E. R.*



**Table 5**

*Vigenere Cipher and the Encryption Equation*

| Key | Key Value (k) | Plaintext | Plaintext Value (p) | Ciphertext Value (c) | c = (p + k) mod 26 | Final Encryption |
|-----|---------------|-----------|---------------------|----------------------|---------------------|------------------|
| D | 3 | M | 11 | 15 | 15 | P |
| I | 8 | A | 0 | 8 | 8 | I |
| S | 18 | T | 19 | 37 | 11 | L |
| C | 2 | H | 7 | 9 | 9 | J |
| R | 17 | I | 8 | 25 | 25 | Z |
| E | 4 | S | 18 | 22 | 22 | W |
| T | 19 | R | 17 | 36 | 10 | K |
| E | 4 | E | 4 | 8 | 8 | I |
| D | 3 | A | 0 | 3 | 3 | D |
| I | 8 | L | 11 | 19 | 19 | T |
| S | 18 | L | 11 | 3 | 3 | D |
| C | 2 | Y | 24 | 0 | 0 | A |
| R | 17 | C | 2 | 19 | 19 | T |
| E | 4 | O | 14 | 18 | 18 | S |
| T | 19 | O | 14 | 7 | 7 | H |
| E | 4 | L | 11 | 15 | 15 | P |

*Note.* Adapted from *The Evolution of Cryptography through Number Theory (p.5)*, by Dawson

Shores, 2021. Copyright 2021 by Dawson Shores.



**Table 6**

*Vigenere Cipher and the Decryption Equation*

| Key | Ciphertext Value (c) | Key | Key Value (k) | Plaintext Value (p) | p = (c + (26 - k)) mod 26 | Plaintext |
|-----|------|-----|------|------|------|------|
| P | 15 | D | 3 | 11 | 11 | M |
| I | 8 | I | 8 | 0 | 0 | A |
| L | 37 | S | 18 | 19 | 19 | T |
| J | 9 | C | 2 | 7 | 7 | H |
| Z | 25 | R | 17 | 8 | 8 | I |
| W | 22 | E | 4 | 18 | 18 | S |
| K | 36 | T | 19 | 17 | 17 | R |
| I | 8 | E | 4 | 4 | 4 | E |
| D | 3 | D | 3 | 0 | 0 | A |
| T | 19 | I | 8 | 11 | 11 | L |
| D | 3 | S | 18 | 11 | 11 | L |
| A | 0 | C | 2 | 24 | 24 | Y |
| T | 19 | R | 17 | 2 | 2 | C |
| S | 18 | E | 4 | 14 | 14 | O |
| H | 7 | T | 19 | 14 | 14 | O |
| P | 15 | E | 4 | 11 | 11 | L |

*Note.* Adapted from *The Evolution of Cryptography through Number Theory (p.5)*, by Dawson

Shores, 2021. Copyright 2021 by Dawson Shores.



**Figure 1**

*The Vigenere Square*

Figure 1: The Vigenère Square

*Note.* From *The vigenere cipher with the TI-83 (p.21), by Michael Hamilton & Bill Yankosky,*

*2004. Copyright 2004 by Michael Hamilton & Bill Yankosky*